\documentclass[%
 reprint,
superscriptaddress,
 amsmath,amssymb,
 aps,
floatfix,
]{revtex4-2}

\usepackage{graphicx}
\usepackage{dcolumn}
\usepackage{bm}
\usepackage{textcomp}
\usepackage{float} 

\begin{document}

\preprint{APS/123-QED}

\title{Direct Imaging of Electrical Switching of Antiferromagnetic \\ N\'eel Order in $\alpha$-Fe$_2$O$_3$ Epitaxial Films}

\author{Egecan Cogulu}
 \email{egecancogulu@nyu.edu}
 \affiliation{Center for Quantum Phenomena, Department of Physics, New York University, USA}
\author{Nahuel N. Statuto}%
 \affiliation{Center for Quantum Phenomena, Department of Physics, New York University, USA}
\author{Yang Cheng}%
 \affiliation{Department of Physics, The Ohio State University, Columbus, OH 43210, USA}
\author{Fengyuan Yang}%
 \affiliation{Department of Physics, The Ohio State University, Columbus, OH 43210, USA}
\author{Rajesh V. Chopdekar}
 \affiliation{Advanced Light Source, Lawrence Berkeley National Laboratories, Berkeley, California 94720, USA}
\author{Hendrik Ohldag}
 \affiliation{Advanced Light Source, Lawrence Berkeley National Laboratories, Berkeley, California 94720, USA}
\author{Andrew D. Kent}%
 \affiliation{Center for Quantum Phenomena, Department of Physics, New York University, USA}

\date{\today}
\begin{abstract}
We report the direct observation of switching of the N\'eel vector of antiferromagnetic (AFM) domains in response to electrical pulses in micron-scale Pt/$\alpha$-Fe$_2$O$_3$ Hall bars using photoemission electron microscopy. Current pulses lead to reversible and repeatable switching, with the current direction determining the final state, consistent with Hall effect experiments that probe only the spatially averaged response. Current pulses also produce irreversible changes in domain structure, in and even outside the current path. In both cases only a fraction of the domains switch in response to pulses. Further, analysis of images taken with different x-ray polarizations shows that the AFM N\'eel order has an out-of-plane component in equilibrium that is important to consider in analyzing the switching data. These results show that---in addition to effects associated with spin-orbit torques from the Pt layer, which can produce reversible switching---changes in AFM order can be induced by purely thermal effects.

\end{abstract}

\maketitle
The electrical control of antiferromagnetic (AFM) order is a topic of great current interest that has been enabled by recent advances in spintronics, specifically the ability to produce spin currents and spin torques based on spin-orbit interactions ~\citep{spintronicsRMP,Miron2011,Liu2012Ta,Hoffmann2013}. AFM states offer advantages compared to those of ferromagnets in that their spin dynamics generally occurs on faster time scales and they are relatively impervious to magnetic fields. They also do not generate macroscopic fields. This makes them both interesting as well as challenging to study, as their microscopic spin structure does not produce strong magnetic signatures. Nonetheless, current-induced switching has been reported in AFM thin films  using electronic transport signatures (e.g. the Hall effect) to infer domain reorientation ~\citep{Wadley2016, Chen2018, Meinert2018,Zhou2018,Bodnar2018,Cheng2020,LuqiaoPRL2020,Moriyama2018,Baldrati2019,Gray2019}. However, recent studies of NiO show that such inferences can be equivocal because electromigration can lead to the same transport response as that ascribed to switching of AFM domains~\citep{Chiang2019,Churikova2020}. It is thus critical to have direct information on the AFM domain response to electrical pulses to advance the understanding of their spin dynamics.

In this work we use x-ray microscopy to directly observe AFM domain structure in Pt/$\alpha$-Fe$_2$O$_3$ Hall bars and their response to electrical pulses. Spatially and element resolved x-ray magnetic linear dichroism (XMLD) photoemission electron microscopy (PEEM) reveals reorientation of N\'eel vector, where the current flow direction repeatably sets and resets the domain orientation. However, domains outside the current path generally switch irreversibly. In both cases, only a small region near and in the Hall cross switches. Further, our analysis of images taken with different x-ray polarizations reveals an out-of-plane component to the N\'eel vector in equilibrium, that is, before current pulses are applied to the sample. These results are consistent with switching of the N\'eel vector between states with in and out of the plane spin components. 

\begin{figure}[t]
  \centering
   \includegraphics[width=1\columnwidth]{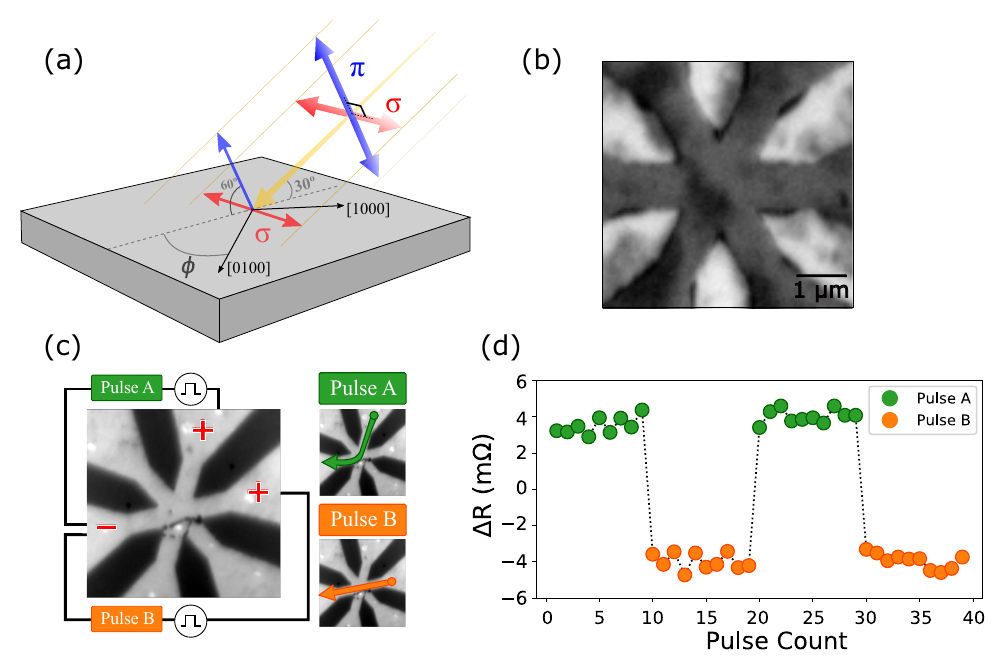}
    \caption{(a) Schematic of the experimental setup showing the two incident X-ray polarizations, $\sigma$ and $\pi$, and their projections on the sample. The sample was rotated by a small azimuthal angle, $\phi$, so that the [1000] and [0100] crystal axes have different projections along $\pi$ and $\sigma$ polarizations. 
    (b) XMLD-PEEM image of the 6-legged Hall cross with AFM domains visible both on bare Fe$_2$O$_3$ regions (lighter regions) and under the 2 nm thick Pt layer (darker regions). 
    (c) Left: PEEM image with a schematic of the circuit. Bright regions correspond to Pt covered areas and dark regions to bare Fe$_2$O$_3$. Pulses were applied from ``+'' to the ground, ``--'', for both pulse generators. Right: Pulse A and Pulse B panels show the current path for each pulse. The dark line at the bottom of the central part of the Hall cross is a cut disconnecting the two bottom leads from the current path. (d) Transverse voltage measurements in response to 4 mA, 10 ms pulses A and B on a sister sample with the bottom leads connected.}
\label{Fig1:xmld_domains}
\end{figure}

Our experiments were designed to detect the AFM domain structure and changes to this structure in response to current pulses. They were carried out at the PEEM3 beamline 11.0.1.1 of the Advanced Light Source at Lawrence Berkeley National Laboratory (LBNL) in the geometry shown in Fig.~\ref{Fig1:xmld_domains}(a)~\citep{DORAN2012340}. X-rays were incident at a 30\textdegree\ angle to the film surface and polarized either in the plane of incidence, labeled $\pi$ polarization (blue arrow), or perpendicular to this plane, $\sigma$ polarization (red arrow). In this configuration, $\sigma$ polarization is completely in the film plane, whereas $\pi$ polarization is at an angle of 60\textdegree\ to the film plane. Moreover, the energy of the incoming photons were resonant with the Fe L$_{2a}$ and L$_{2b}$ edges making their absorption dependent on the magnetic order (see Fig. S1 in supplementary materials). 
The XMLD response thus allows characterization of the AFM order both in---with $\sigma$ polarization---and out of the film plane, with $\pi$ polarization x-rays. The XMLD contrast is typically a few  percent, which is easily detectable with PEEM. 

The experiments were performed on 30 nm thick epitaxial c-axis oriented $\alpha$-Fe$_2$O$_3$ capped with 2 nm of Pt~\cite{Cheng2020}. The Pt is patterned so that the arms of the Hall bar are parallel to the three in-plane magnetic easy axes of Fe$_2$O$_3$~\cite{Cheng2020}. Figure\ \ref{Fig1:xmld_domains}(b) shows a XMLD-PEEM image of the sample. The darker regions are covered with the thin Pt layer and the brighter regions are where the Pt is etched, exposing the $\alpha$-Fe$_2$O$_3$ surface. AFM domains are observed in both regions. Their boundaries are gray to black contrast changes in the Pt covered regions and the white to gray transitions in the $\alpha$-Fe$_2$O$_3$ exposed regions. The lateral scale of the domains is $\sim 1\; \mu$m and domains extend between the two regions. 

To characterize the orientation of the N\'eel vector, XMLD images were acquired as a function of the sample orientation, $\phi$ in Fig. \ref{Fig1:xmld_domains}(a). XMLD contrast depends on the {\em magnitude} of the projection of the x-ray polarization $\hat{p}$ on the N\'eel vector $\hat{n}$, $(\hat{p}\cdot \hat{n})^2$. Therefore, for in-plane oriented spins, XMLD images are invariant under a 180$^\circ$ sample rotation for both $\sigma$ and $\pi$ polarizations. Based on the spin structure and studies of $\alpha$-Fe$_2$O$_3$ thin films, we expected the N\'eel vector to be oriented along in-plane easy axis directions~\cite{Morrish1995,Cheng2020,LuqiaoPRL2020} and thus that both sets of images would be invariant under 180$^\circ$ sample rotation.

However, surprisingly, our results show that for $\pi$ polarization, the image contrast changes significantly on 180$^\circ$ rotation. The $\sigma$ polarization images return to the same contrast levels on rotation, as expected (see Fig. S2 in supplementary materials). This indicates that, in addition to having different in-plane projections, the AFM N\'eel vector at the sample surface is appreciably canted out of the film plane, which may reflect an interface perpendicular magnetic anisotropy.

\begin{figure}[t]
  \centering
   \includegraphics[width=1\columnwidth]{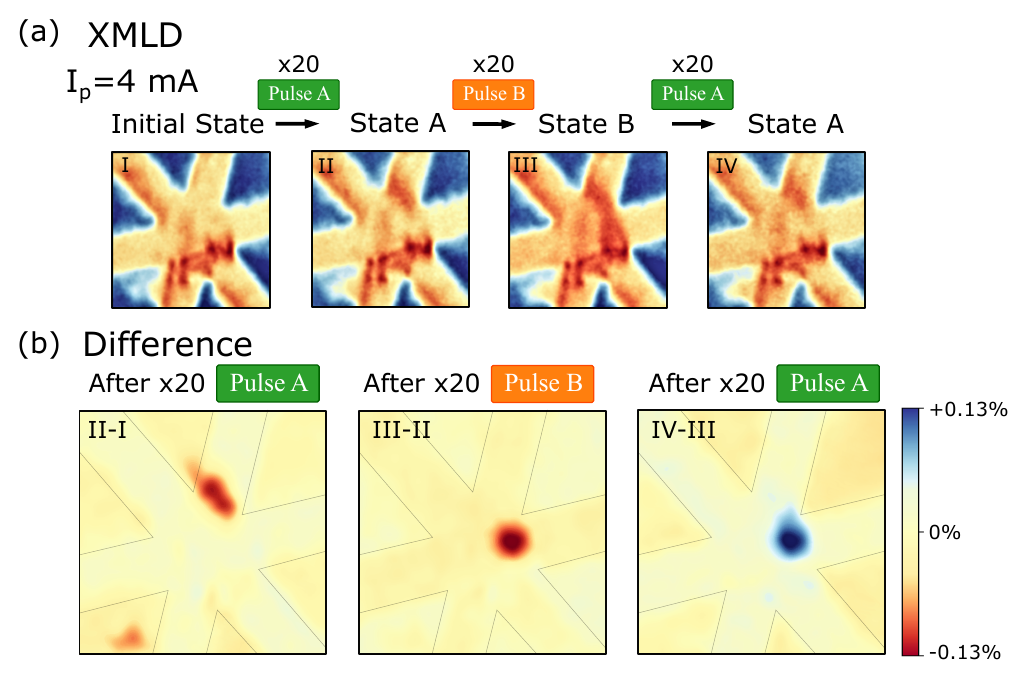}
    \caption{AFM domain changes in response to current pulses. (a) XMLD images showing a $4\times 4\;\mu$m$^2$ field of view. Image I shows the sample before current is applied; Images II, III and IV are images taken after the application of 20 current pulses of type A, type B and then type A, respectively. The pulse amplitude is 4 mA for all the images in the sequence. (b) Differences of XMLD images in panel (a) to highlight the pulse induced changes to AFM domain structure. The color scale in (b) is expanded to highlight changes.}
\label{Fig2:xmld_4mA}
\end{figure}

The effect of electrical pulses on the AFM domain structure were studied by applying currents to the Hall bar in situ, with the sample in the PEEM microscope. Prior electrical studies have shown that the current flow direction leads to different transverse (Hall) voltages that were associated with changes in the AFM domain structure~\cite{Cheng2020,LuqiaoPRL2020}. Thus we configured the sample to enable current pulses to be applied in two different directions. Figure~\ref{Fig1:xmld_domains}(c) shows the sample layout with connections to two separate pulse generators, labeled A and B. The images on the right of Fig.~\ref{Fig1:xmld_domains}(c) show the current directions for A and B pulses.

Samples were also characterized electrically prior to XMLD PEEM experiments. Current pulses were applied
and the transverse voltage was measured. The data shown in Fig.~\ref{Fig1:xmld_domains}(d) is for 4 mA amplitude 10 ms duration pulses. After each pulse a small sensing current 100 $\mu$A ($\ll 4$ mA) was used to measure the transverse voltage. The data points are color coded according to the pulse type. As reported previously, there is a step change in voltage when going from A to B type pulses as well as going from B to A type pulses~\cite{Cheng2020}. Further, subsequent pulses of the same type do not change the transverse voltage.

Figure~\ref{Fig2:xmld_4mA}(a) shows a sequence of four XMLD images taken before and after applying current pulses.
They show the difference in the electron yield for $\sigma$ and $\pi$ polarization to maximize the signal originating from AFM order.
Blue regions are bare Fe$_2$O$_3$, and red/orange regions correspond to the Pt leads. The field of view is $4 \times 4\;\mu$m$^2$. Image I shows the initial domain configuration before pulses were applied. Images II, III and IV show the states after sending a sequence of 20 of the same pulse type, first pulse A, then B, and then A again. The labels ``State A'' and ``State B'' above the images correspond to the preceding pulse type. Changes in AFM order correspond to color changes between subsequent images. 
\begin{figure}[t]
  \centering
   \includegraphics[width=1 \columnwidth]{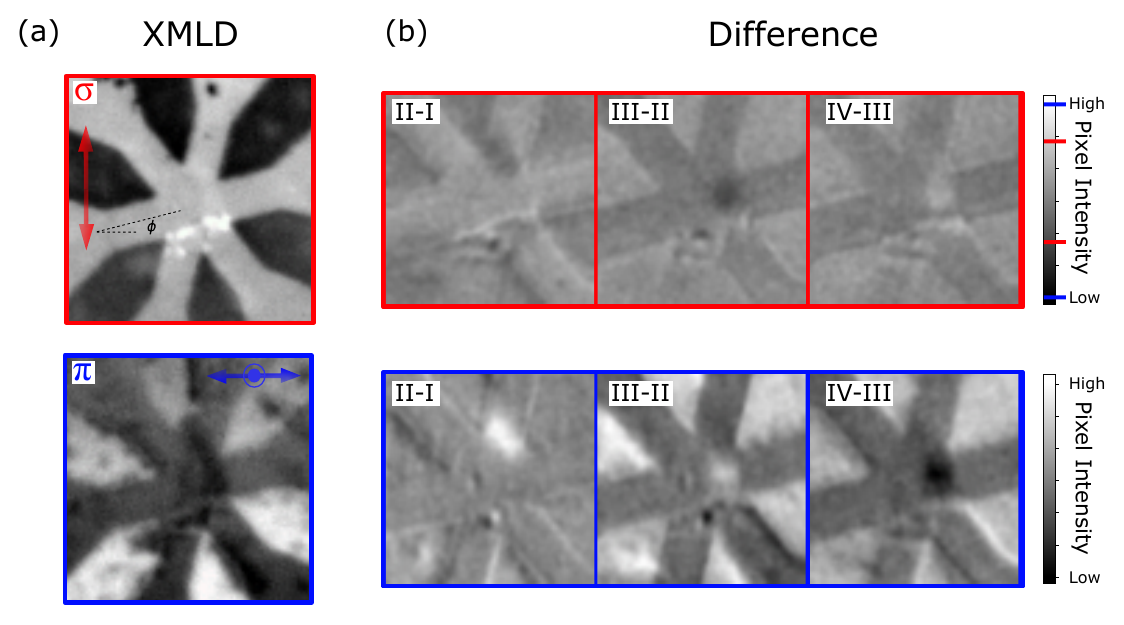}
    \caption{(a) $\sigma$ and $\pi$ projections of Image III in Fig.\ \ref{Fig2:xmld_4mA}a. Bright regions indicate that the AFM order is aligned perpendicular to the polarization direction whereas dark regions indicate that the AFM order aligns parallel to the polarization direction. (b) $\sigma$ (top panel) and $\pi$ (bottom panel) projections of the difference images in Fig.\ \ref{Fig2:xmld_4mA}b. The same scale bar is used on both panels to highlight the larger contrast in the $\pi$ polarization images. The marks on the $\sigma$ color scale bar show the maximum and minimum values for $\sigma$ (red marks) and $\pi$ (blue marks) polarization difference images.
    }
\label{Fig3:Polarizations}
\end{figure}
For example, in Image II there is added contrast in the pulse A positive lead after 20 type A pulses and in Image III a domain to the right of the center of the Hall cross has reoriented after 20 type B pulses. While in Image IV, after A pulses, this domain has returned to its initial configuration. 

To bring out the changes in AFM domain structure we show difference images in Fig.~\ref{Fig2:xmld_4mA}(b). In these images blue regions represent positive contrast changes and red regions negative contrast changes. The color map is chosen to highlight the current-induced changes; the black lines indicate the boundaries of the Pt leads. The II-I Image clearly shows the domain that has reoriented in response to the first set of A pulses is in one of the current leads to the Hall cross. Image III-II shows the domain that has switched in the Hall cross area in response to B pulses. After another set of A pulses, the contrast returns to that after the first set of A pulses, ``State A,'' the state shown in Fig.~\ref{Fig2:xmld_4mA}(a)II. 

From these results, we identify two types of domain changes in response to current pulses: 1) reversible changes, regions that go back and forth between State A and State B after the corresponding pulses; and 2) irreversible regions in which the domain configuration does not return to its initial state.  The reversible changes occur in regions in which the current density is large and its flow direction changes significantly between that of pulse A and B (see Fig. S4 in supplementary materials for the current flow directions). In both cases, it is also clear that the N\'eel vector of only a fraction of the domains in the current path change their orientation in response to pulses.

To determine the changes in the orientation of the N\'eel vector induced by current pulses we now analyze data from the $\sigma$ and $\pi$ polarization data separately. Figure\ \ref{Fig3:Polarizations}(a) shows the $\sigma$ and $\pi$ polarization data that are used to form Image III in Fig.~\ref{Fig2:xmld_4mA}(a).  

The $\sigma$ polarization is in the film plane and thus provides sensitivity to changes in the projection of the N\'eel vector in this plane, while the $\pi$ polarization has both in and out of the film plane projections. In these images, bright contrast in a region indicates that the AFM order is mostly aligned perpendicular to the polarization direction, whereas dark contrast indicates that the AFM order aligns mostly parallel to the polarization direction. Although the contrast associated with the switched AFM domain is visible in both polarization images, the contrast is much stronger in the $\pi$-polarization images. Moreover, dark contrast of the AFM switched region in the $\pi$-polarization image indicates that the projection of the N\'eel vector on the $\pi$ polarization has changed significantly in response to the current pulse.

We now separately analyze the polarization dependence of the difference images in Fig.\ \ref{Fig2:xmld_4mA}(b). The results are shown in Fig.\ \ref{Fig3:Polarizations}(b). The top row of images (red bordered images) shows the $\sigma$ differences images that highlight changes to the in-plane AFM order. The lower row of blue bordered images show the $\pi$ difference images that are sensitive to the projection on the $\pi$ polarization direction. All six differences images are on the same color scale to be able to compare the changes in contrast. The contrast change in $\pi$-polarization images is approximately twice as large as that of the $\sigma$-polarization images. This indicates that the projection of the N\'eel vector on the $\pi$ polarization changes more than its projection on the $\sigma$ polarization direction. 

\begin{figure}[t]
  \centering
   \includegraphics[width=1 \columnwidth]{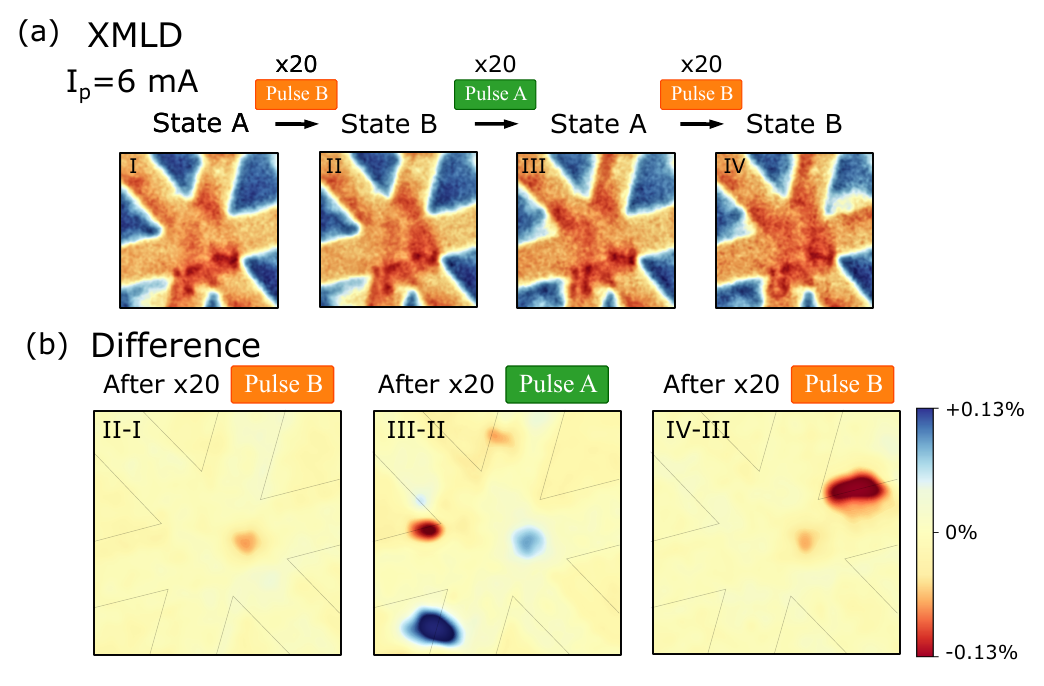}
    \caption{AFM domain response to larger amplitude current pulses. (I) (a) I. State A, initial sample state. Followed by images after II. 20 $\times$ B pulses,  III;  20 $\times$ A pulses, IV.  20 $\times$ B pulses. In each case the current amplitude was 6 mA. (b) Differences of XMLD images in panel (a) highlight the pulse induced changes, with an expanded color scale, the same as the that of Fig.~\ref{Fig2:xmld_4mA}(b).}
\label{Fig4:high_current}
\end{figure}

We further studied the effect of higher currents. We repeated the same pulse sequences, as in Fig. \ref{Fig2:xmld_4mA}, but this time with a 50\% higher current pulse amplitude, 6 mA. Figure \ref{Fig4:high_current}(a) shows XMLD images taken before and after applying the current pulses. Starting from State A (Fig.~\ref{Fig2:xmld_4mA}(a)II) and applying 20 current pulses of B, then A and then B again. Figure~\ref{Fig4:high_current}(b) shows the differences between images, again to highlight the change in domain structure associated with the current pulses. 
Figure~\ref{Fig4:high_current}(b) shows the differences between the images, again to highlight the change in domain structure associated with the current pulses.

Besides the fact that the reversible domain reorientation is still observable, the light blue dot in Fig. \ref{Fig4:high_current}(b), at this larger current there are additional changes. 
In difference panels (III-II) and (IV-III)  blue and red regions represent new AFM domains switched by the application of the pulse A and that most of them are irreversible switches. In contrast to Fig.~\ref{Fig2:xmld_4mA}, where almost all the switched AFM domains were in the current path, Fig. \ref{Fig4:high_current}(b) shows switched regions outside of the current path. In fact, the blue region on panel (III-II) is situated well outside the region that experiences the current.

\begin{figure}[t]
  \centering
   \includegraphics[width=0.8 \columnwidth]{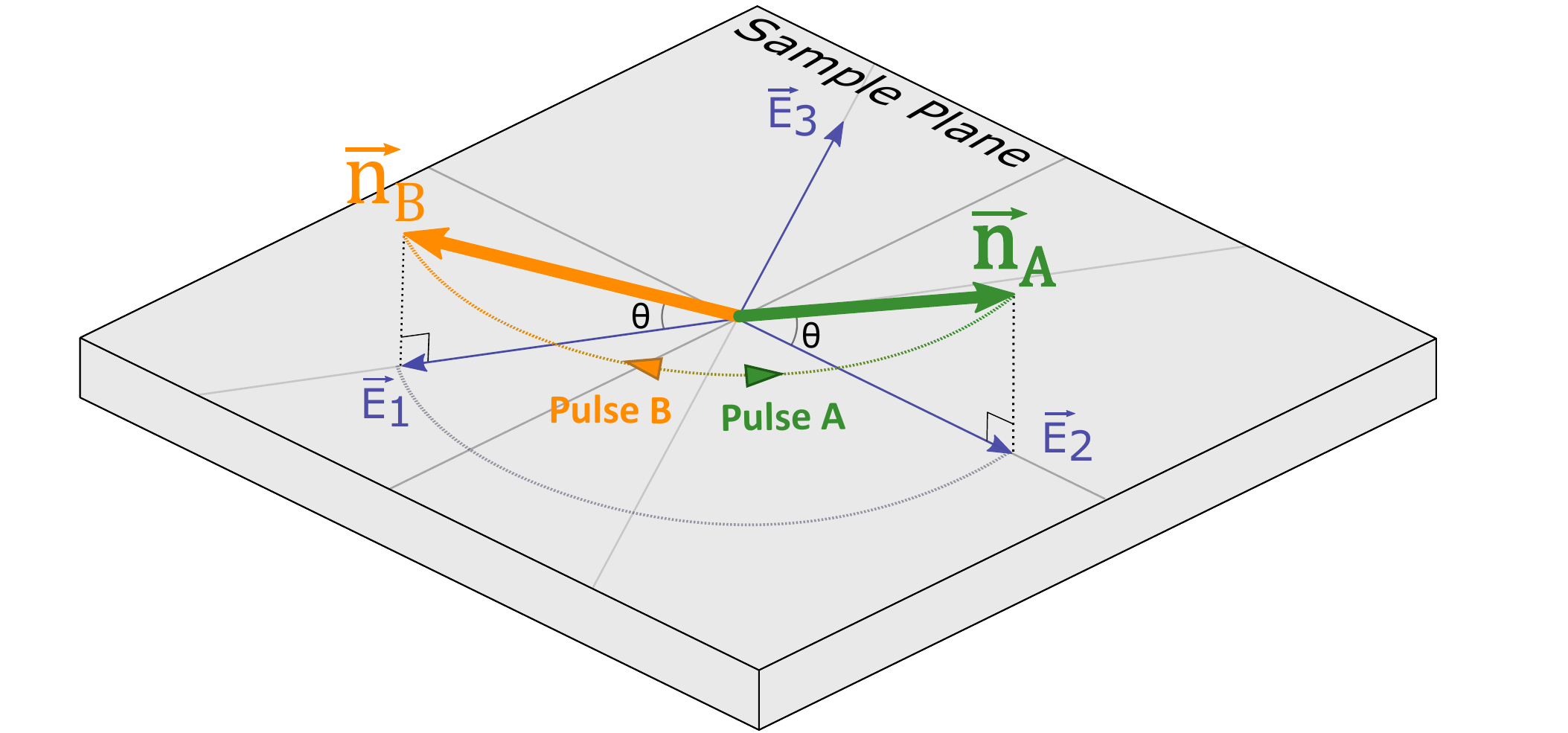}
    \caption{Schematic of the switching of the N\'eel vector between easy axis directions that cant out of the film plane, which is consistent with both the XMLD-PEEM results and the Hall response. $\vec{n}_{A}$ and  $\vec{n}_{B}$ are the orientations of the N\'eel vector after Pulse A and Pulse B respectively. The dotted lines with arrows shows the switching direction of the N\'eel vector in response to the pulses.}
\label{Fig5:3Dswitch}
\end{figure}

Previous work on current-induced switching in $\alpha$-Fe$_2$O$_3$ assumes that the N\'eel vector lies and rotates completely in the film~\cite{Cheng2020,LuqiaoPRL2020}.
However, our sample characterization and polarization dependent analysis show that this is not the case. The fact that the contrast of the AFM domains changes in the $\pi$ projection images after a 180\textdegree\ rotation combined with the fact that we are observing more contrast in the $\pi$ polarization images indicate that N\'eel vector has an equilibrium out-of-plane component. Our results show that the current-induced switching is between easy axes that are canted out of the film plane as indicated in Fig.~\ref{Fig5:3Dswitch}.
 
Now in the light of these observations, we turn to the question of the potential switching mechanisms. Two main mechanisms were used to explain the observed magnetoresistance, spin-orbit-torques (SOT) and thermally induced magnetostriction. For the spin-orbit-torque case, the spin accumulation generated in the Pt acts as an effective magnetic field and exerts a damping-like SOT on the N\'eel order to align it with the current direction. Whereas for thermally induced magnetostriction, compressive normal stresses induced by Joule heating changes the anisotropy energy through magnetostriction, thus, also favoring alignment of the N\'eel vector with the applied current. In easy-plane antiferromagnets with large magnetostrictive coefficients such as $\alpha$-Fe$_2$O$_3$ and NiO, this can outweigh damping-like torques \cite{LuqiaoPRL2020,meer2020direct}. Our finite element simulations (see supplementary materials) show that both the current density levels ($\sim2\times10^8$ A/cm$^2$) and thermally induced compressive normal stresses are above their respective required thresholds ($\sim$50 MPa) to induce switching. This suggests that both mechanisms potentially contribute to the switching. The fact that we can see changes in the N\'eel vector outside of the current path demonstrates that thermally induced magnetostriction alone is enough to produce switching, since there cannot be any SOT effects where there is no current. However, we also see all of the reversible domain changes in the current flow path, i.e. where SOT is present. 

To conclude, we directly imaged repeatable current-induced switching of antiferromagnetic moments in $\alpha$-Fe$_2$O$_3$/Pt bilayers with XMLD-PEEM. We observed that only a fraction of the domains reorient, consistent with electrical measurements---which shows that the Hall effect response is less than that expected for complete switching in the Hall cross region. We further identified two types of response that tend to occur in different regions: reversible switches in the Hall cross area and irreversible switches in and outside the Hall cross. While both SOT and thermally induced magnetostriction potentially contribute to switching, our experiments with high current density pulses demonstrate that thermal effects alone can induce changes in the N\'eel order, by showing there are irreversible changes to the AFM domain structure outside the electrical path. To gain further insights into the switching mechanisms, the impacts of SOT and thermal effects on AFM order needs to be separated, such as by electrically (but not thermally) isolating the metal layers from the $\alpha$-Fe$_2$O$_3$, or using light metals (e.g. Al) instead of Pt, which will be the topics of future studies.

\section*{Acknowledgements}

This research was supported by the Air Force Office of Scientific Research under Grant FA9550- 19-1-0307. The nanostructures were realized at the Advanced Science Research Center NanoFabrication Facility of the Graduate Center at the City University of New York. This research also used resources of the Advanced Light Source, a U.S. DOE Office of Science User Facility under contract no. DE-AC02-05CH11231. Beamline 11.0.1.1 was used for XMLD-PEEM imaging and Beamline 6.3.1 was used for magnetic spectroscopy.


\clearpage


\begin{thebibliography}{21}%
\makeatletter
\providecommand \@ifxundefined [1]{%
 \@ifx{#1\undefined}
}%
\providecommand \@ifnum [1]{%
 \ifnum #1\expandafter \@firstoftwo
 \else \expandafter \@secondoftwo
 \fi
}%
\providecommand \@ifx [1]{%
 \ifx #1\expandafter \@firstoftwo
 \else \expandafter \@secondoftwo
 \fi
}%
\providecommand \natexlab [1]{#1}%
\providecommand \enquote  [1]{``#1''}%
\providecommand \bibnamefont  [1]{#1}%
\providecommand \bibfnamefont [1]{#1}%
\providecommand \citenamefont [1]{#1}%
\providecommand \href@noop [0]{\@secondoftwo}%
\providecommand \href [0]{\begingroup \@sanitize@url \@href}%
\providecommand \@href[1]{\@@startlink{#1}\@@href}%
\providecommand \@@href[1]{\endgroup#1\@@endlink}%
\providecommand \@sanitize@url [0]{\catcode `\\12\catcode `\$12\catcode
  `\&12\catcode `\#12\catcode `\^12\catcode `\_12\catcode `\%12\relax}%
\providecommand \@@startlink[1]{}%
\providecommand \@@endlink[0]{}%
\providecommand \url  [0]{\begingroup\@sanitize@url \@url }%
\providecommand \@url [1]{\endgroup\@href {#1}{\urlprefix }}%
\providecommand \urlprefix  [0]{URL }%
\providecommand \Eprint [0]{\href }%
\providecommand \doibase [0]{http://dx.doi.org/}%
\providecommand \selectlanguage [0]{\@gobble}%
\providecommand \bibinfo  [0]{\@secondoftwo}%
\providecommand \bibfield  [0]{\@secondoftwo}%
\providecommand \translation [1]{[#1]}%
\providecommand \BibitemOpen [0]{}%
\providecommand \bibitemStop [0]{}%
\providecommand \bibitemNoStop [0]{.\EOS\space}%
\providecommand \EOS [0]{\spacefactor3000\relax}%
\providecommand \BibitemShut  [1]{\csname bibitem#1\endcsname}%
\let\auto@bib@innerbib\@empty
\bibitem [{\citenamefont {Baltz}\ \emph {et~al.}(2018)\citenamefont {Baltz},
  \citenamefont {Manchon}, \citenamefont {Tsoi}, \citenamefont {Moriyama},
  \citenamefont {Ono},\ and\ \citenamefont {Tserkovnyak}}]{spintronicsRMP}%
  \BibitemOpen
  \bibfield  {author} {\bibinfo {author} {\bibfnamefont {V.}~\bibnamefont
  {Baltz}}, \bibinfo {author} {\bibfnamefont {A.}~\bibnamefont {Manchon}},
  \bibinfo {author} {\bibfnamefont {M.}~\bibnamefont {Tsoi}}, \bibinfo {author}
  {\bibfnamefont {T.}~\bibnamefont {Moriyama}}, \bibinfo {author}
  {\bibfnamefont {T.}~\bibnamefont {Ono}}, \ and\ \bibinfo {author}
  {\bibfnamefont {Y.}~\bibnamefont {Tserkovnyak}},\ }\bibfield  {title}
  {\enquote {\bibinfo {title} {Antiferromagnetic spintronics},}\ }\href
  {\doibase 10.1103/RevModPhys.90.015005} {\bibfield  {journal} {\bibinfo
  {journal} {Rev. Mod. Phys.}\ }\textbf {\bibinfo {volume} {90}},\ \bibinfo
  {pages} {015005} (\bibinfo {year} {2018})}\BibitemShut {NoStop}%
\bibitem [{\citenamefont {Mihai~Miron}\ \emph {et~al.}({2011})\citenamefont
  {Mihai~Miron}, \citenamefont {Garello}, \citenamefont {Gaudin}, \citenamefont
  {Zermatten}, \citenamefont {Costache}, \citenamefont {Auffret}, \citenamefont
  {Bandiera}, \citenamefont {Rodmacq}, \citenamefont {Schuhl},\ and\
  \citenamefont {Gambardella}}]{Miron2011}
  \BibitemOpen
  \bibfield  {author} {\bibinfo {author} {\bibfnamefont {Ioan}\ \bibnamefont
  {Mihai~Miron}}, \bibinfo {author} {\bibfnamefont {Kevin}\ \bibnamefont
  {Garello}}, \bibinfo {author} {\bibfnamefont {Gilles}\ \bibnamefont
  {Gaudin}}, \bibinfo {author} {\bibfnamefont {Pierre-Jean}\ \bibnamefont
  {Zermatten}}, \bibinfo {author} {\bibfnamefont {Marius~V.}\ \bibnamefont
  {Costache}}, \bibinfo {author} {\bibfnamefont {Stephane}\ \bibnamefont
  {Auffret}}, \bibinfo {author} {\bibfnamefont {Sebastien}\ \bibnamefont
  {Bandiera}}, \bibinfo {author} {\bibfnamefont {Bernard}\ \bibnamefont
  {Rodmacq}}, \bibinfo {author} {\bibfnamefont {Alain}\ \bibnamefont {Schuhl}},
  \ and\ \bibinfo {author} {\bibfnamefont {Pietro}\ \bibnamefont
  {Gambardella}},\ }\bibfield  {title} {\enquote {\bibinfo {title}
  {{Perpendicular switching of a single ferromagnetic layer induced by in-plane
  current injection}},}\ }\href {\doibase {10.1038/nature10309}} {\bibfield
  {journal} {\bibinfo  {journal} {{Nature}}\ }\textbf {\bibinfo {volume}
  {{476}}},\ \bibinfo {pages} {{189}} (\bibinfo {year} {{2011}})}\BibitemShut
  {NoStop}%
\bibitem [{\citenamefont {Liu}\ \emph {et~al.}({2012})\citenamefont {Liu},
  \citenamefont {Pai}, \citenamefont {Li}, \citenamefont {Tseng}, \citenamefont
  {Ralph},\ and\ \citenamefont {Buhrman}}]{Liu2012Ta}
  \BibitemOpen
  \bibfield  {author} {\bibinfo {author} {\bibfnamefont {Luqiao}\ \bibnamefont
  {Liu}}, \bibinfo {author} {\bibfnamefont {Chi-Feng}\ \bibnamefont {Pai}},
  \bibinfo {author} {\bibfnamefont {Y.}~\bibnamefont {Li}}, \bibinfo {author}
  {\bibfnamefont {H.~W.}\ \bibnamefont {Tseng}}, \bibinfo {author}
  {\bibfnamefont {D.~C.}\ \bibnamefont {Ralph}}, \ and\ \bibinfo {author}
  {\bibfnamefont {R.~A.}\ \bibnamefont {Buhrman}},\ }\bibfield  {title}
  {\enquote {\bibinfo {title} {{Spin-Torque Switching with the Giant Spin Hall
  Effect of Tantalum}},}\ }\href {\doibase {10.1126/science.1218197}}
  {\bibfield  {journal} {\bibinfo  {journal} {{Science}}\ }\textbf {\bibinfo
  {volume} {{336}}},\ \bibinfo {pages} {{555--558}} (\bibinfo {year}
  {{2012}})}\BibitemShut {NoStop}%
\bibitem [{\citenamefont {Hoffmann}({2013})}]{Hoffmann2013}%
  \BibitemOpen
  \bibfield  {author} {\bibinfo {author} {\bibfnamefont {Axel}\ \bibnamefont
  {Hoffmann}},\ }\bibfield  {title} {\enquote {\bibinfo {title} {{Spin Hall
  Effects in Metals}},}\ }\href {\doibase {10.1109/TMAG.2013.2262947}}
  {\bibfield  {journal} {\bibinfo  {journal} {{IEEE Transactions on
  Magnetics}}\ }\textbf {\bibinfo {volume} {{49}}},\ \bibinfo {pages}
  {{5172--5193}} (\bibinfo {year} {{2013}})}\BibitemShut {NoStop}%
\bibitem [{\citenamefont {Wadley}\ \emph {et~al.}(2016)\citenamefont {Wadley},
  \citenamefont {Howells}, \citenamefont {{\v{Z}}elezny}, \citenamefont
  {Andrews}, \citenamefont {Hills}, \citenamefont {Campion}, \citenamefont
  {Novak}, \citenamefont {Olejnik}, \citenamefont {Maccherozzi}, \citenamefont
  {Dhesi}, \citenamefont {Martin}, \citenamefont {Wagner}, \citenamefont
  {Wunderlich}, \citenamefont {Freimuth}, \citenamefont {Mokrousov},
  \citenamefont {Kune}, \citenamefont {Chauhan}, \citenamefont {Grzybowski},
  \citenamefont {Rushforth}, \citenamefont {Edmonds}, \citenamefont
  {Gallagher},\ and\ \citenamefont {Jungwirth}}]{Wadley2016}%
  \BibitemOpen
  \bibfield  {author} {\bibinfo {author} {\bibfnamefont {P.}~\bibnamefont
  {Wadley}}, \bibinfo {author} {\bibfnamefont {B.}~\bibnamefont {Howells}},
  \bibinfo {author} {\bibfnamefont {J.}~\bibnamefont {{\v{Z}}elezny}}, \bibinfo
  {author} {\bibfnamefont {C.}~\bibnamefont {Andrews}}, \bibinfo {author}
  {\bibfnamefont {V.}~\bibnamefont {Hills}}, \bibinfo {author} {\bibfnamefont
  {R.~P.}\ \bibnamefont {Campion}}, \bibinfo {author} {\bibfnamefont
  {V.}~\bibnamefont {Novak}}, \bibinfo {author} {\bibfnamefont
  {K.}~\bibnamefont {Olejnik}}, \bibinfo {author} {\bibfnamefont
  {F.}~\bibnamefont {Maccherozzi}}, \bibinfo {author} {\bibfnamefont {S.~S.}\
  \bibnamefont {Dhesi}}, \bibinfo {author} {\bibfnamefont {S.~Y.}\ \bibnamefont
  {Martin}}, \bibinfo {author} {\bibfnamefont {T.}~\bibnamefont {Wagner}},
  \bibinfo {author} {\bibfnamefont {J.}~\bibnamefont {Wunderlich}}, \bibinfo
  {author} {\bibfnamefont {F.}~\bibnamefont {Freimuth}}, \bibinfo {author}
  {\bibfnamefont {Y.}~\bibnamefont {Mokrousov}}, \bibinfo {author}
  {\bibfnamefont {J.}~\bibnamefont {Kune}}, \bibinfo {author} {\bibfnamefont
  {J.~S.}\ \bibnamefont {Chauhan}}, \bibinfo {author} {\bibfnamefont {M.~J.}\
  \bibnamefont {Grzybowski}}, \bibinfo {author} {\bibfnamefont {A.~W.}\
  \bibnamefont {Rushforth}}, \bibinfo {author} {\bibfnamefont {K.~W.}\
  \bibnamefont {Edmonds}}, \bibinfo {author} {\bibfnamefont {B.~L.}\
  \bibnamefont {Gallagher}}, \ and\ \bibinfo {author} {\bibfnamefont
  {T.}~\bibnamefont {Jungwirth}},\ }\bibfield  {title} {\enquote {\bibinfo
  {title} {Electrical switching of an antiferromagnet},}\ }\href {\doibase
  10.1126/science.aab1031} {\bibfield  {journal} {\bibinfo  {journal}
  {Science}\ }\textbf {\bibinfo {volume} {351}},\ \bibinfo {pages} {587}
  (\bibinfo {year} {2016})}\BibitemShut {NoStop}%
\bibitem [{\citenamefont {Chen}\ \emph {et~al.}(2018)\citenamefont {Chen},
  \citenamefont {Zarzuela}, \citenamefont {Zhang}, \citenamefont {Song},
  \citenamefont {Zhou}, \citenamefont {Shi}, \citenamefont {Li}, \citenamefont
  {Zhou}, \citenamefont {Jiang}, \citenamefont {Pan},\ and\ \citenamefont
  {Tserkovnyak}}]{Chen2018}%
  \BibitemOpen
  \bibfield  {author} {\bibinfo {author} {\bibfnamefont {X.~Z.}\ \bibnamefont
  {Chen}}, \bibinfo {author} {\bibfnamefont {R.}~\bibnamefont {Zarzuela}},
  \bibinfo {author} {\bibfnamefont {J.}~\bibnamefont {Zhang}}, \bibinfo
  {author} {\bibfnamefont {C.}~\bibnamefont {Song}}, \bibinfo {author}
  {\bibfnamefont {X.~F.}\ \bibnamefont {Zhou}}, \bibinfo {author}
  {\bibfnamefont {G.~Y.}\ \bibnamefont {Shi}}, \bibinfo {author} {\bibfnamefont
  {F.}~\bibnamefont {Li}}, \bibinfo {author} {\bibfnamefont {H.~A.}\
  \bibnamefont {Zhou}}, \bibinfo {author} {\bibfnamefont {W.~J.}\ \bibnamefont
  {Jiang}}, \bibinfo {author} {\bibfnamefont {F.}~\bibnamefont {Pan}}, \ and\
  \bibinfo {author} {\bibfnamefont {Y.}~\bibnamefont {Tserkovnyak}},\
  }\bibfield  {title} {\enquote {\bibinfo {title} {Antidamping-torque-induced
  switching in biaxial antiferromagnetic insulators},}\ }\href {\doibase
  10.1103/PhysRevLett.120.207204} {\bibfield  {journal} {\bibinfo  {journal}
  {Phys. Rev. Lett.}\ }\textbf {\bibinfo {volume} {120}},\ \bibinfo {pages}
  {207204} (\bibinfo {year} {2018})}\BibitemShut {NoStop}%
\bibitem [{\citenamefont {Meinert}\ \emph {et~al.}(2018)\citenamefont
  {Meinert}, \citenamefont {Graulich},\ and\ \citenamefont
  {Matalla-Wagner}}]{Meinert2018}%
  \BibitemOpen
  \bibfield  {author} {\bibinfo {author} {\bibfnamefont {Markus}\ \bibnamefont
  {Meinert}}, \bibinfo {author} {\bibfnamefont {Dominik}\ \bibnamefont
  {Graulich}}, \ and\ \bibinfo {author} {\bibfnamefont {Tristan}\ \bibnamefont
  {Matalla-Wagner}},\ }\bibfield  {title} {\enquote {\bibinfo {title}
  {{Electrical Switching of Antiferromagnetic ${\mathrm{Mn}}_{2}\mathrm{Au}$
  and the Role of Thermal Activation}},}\ }\href {\doibase
  10.1103/PhysRevApplied.9.064040} {\bibfield  {journal} {\bibinfo  {journal}
  {Phys. Rev. Applied}\ }\textbf {\bibinfo {volume} {9}},\ \bibinfo {pages}
  {064040} (\bibinfo {year} {2018})}\BibitemShut {NoStop}%
\bibitem [{\citenamefont {Zhou}\ \emph {et~al.}(2018)\citenamefont {Zhou},
  \citenamefont {Zhang}, \citenamefont {Li}, \citenamefont {Chen},
  \citenamefont {Shi}, \citenamefont {Tan}, \citenamefont {Gu}, \citenamefont
  {Saleem}, \citenamefont {Wu}, \citenamefont {Pan},\ and\ \citenamefont
  {Song}}]{Zhou2018}%
  \BibitemOpen
  \bibfield  {author} {\bibinfo {author} {\bibfnamefont {X.~F.}\ \bibnamefont
  {Zhou}}, \bibinfo {author} {\bibfnamefont {J.}~\bibnamefont {Zhang}},
  \bibinfo {author} {\bibfnamefont {F.}~\bibnamefont {Li}}, \bibinfo {author}
  {\bibfnamefont {X.~Z.}\ \bibnamefont {Chen}}, \bibinfo {author}
  {\bibfnamefont {G.~Y.}\ \bibnamefont {Shi}}, \bibinfo {author} {\bibfnamefont
  {Y.~Z.}\ \bibnamefont {Tan}}, \bibinfo {author} {\bibfnamefont {Y.~D.}\
  \bibnamefont {Gu}}, \bibinfo {author} {\bibfnamefont {M.~S.}\ \bibnamefont
  {Saleem}}, \bibinfo {author} {\bibfnamefont {H.~Q.}\ \bibnamefont {Wu}},
  \bibinfo {author} {\bibfnamefont {F.}~\bibnamefont {Pan}}, \ and\ \bibinfo
  {author} {\bibfnamefont {C.}~\bibnamefont {Song}},\ }\bibfield  {title}
  {\enquote {\bibinfo {title} {{Strong Orientation-Dependent Spin-Orbit Torque
  in Thin Films of the Antiferromagnet Mn$_2$Au}},}\ }\href {\doibase
  10.1103/PhysRevApplied.9.054028} {\bibfield  {journal} {\bibinfo  {journal}
  {Phys. Rev. Applied}\ }\textbf {\bibinfo {volume} {9}},\ \bibinfo {pages}
  {054028} (\bibinfo {year} {2018})}\BibitemShut {NoStop}%
\bibitem [{\citenamefont {Bodnar}\ \emph {et~al.}(2018)\citenamefont {Bodnar},
  \citenamefont {{\v{S}}mejkal}, \citenamefont {Turek}, \citenamefont
  {Jungwirth}, \citenamefont {Gomonay}, \citenamefont {Sinova}, \citenamefont
  {Sapozhnik}, \citenamefont {Elmers}, \citenamefont {Kl\"{a}ui},\ and\
  \citenamefont {Jourdan}}]{Bodnar2018}%
  \BibitemOpen
  \bibfield  {author} {\bibinfo {author} {\bibfnamefont {S.~Yu.}\ \bibnamefont
  {Bodnar}}, \bibinfo {author} {\bibfnamefont {L.}~\bibnamefont
  {{\v{S}}mejkal}}, \bibinfo {author} {\bibfnamefont {I.}~\bibnamefont
  {Turek}}, \bibinfo {author} {\bibfnamefont {T.}~\bibnamefont {Jungwirth}},
  \bibinfo {author} {\bibfnamefont {O.}~\bibnamefont {Gomonay}}, \bibinfo
  {author} {\bibfnamefont {J.}~\bibnamefont {Sinova}}, \bibinfo {author}
  {\bibfnamefont {A.~A.}\ \bibnamefont {Sapozhnik}}, \bibinfo {author}
  {\bibfnamefont {H.-J.}\ \bibnamefont {Elmers}}, \bibinfo {author}
  {\bibfnamefont {M.}~\bibnamefont {Kl\"{a}ui}}, \ and\ \bibinfo {author}
  {\bibfnamefont {M.}~\bibnamefont {Jourdan}},\ }\bibfield  {title} {\enquote
  {\bibinfo {title} {{Writing and reading antiferromagnetic Mn$_2$Au by
  N{\'{e}}el spin-orbit torques and large anisotropic magnetoresistance}},}\
  }\href {https://doi.org/10.1038/s41467-017-02780-x} {\bibfield  {journal}
  {\bibinfo  {journal} {Nature Communications}\ }\textbf {\bibinfo {volume}
  {9}} (\bibinfo {year} {2018})}\BibitemShut {NoStop}%
\bibitem [{\citenamefont {Cheng}\ \emph {et~al.}(2020)\citenamefont {Cheng},
  \citenamefont {Yu}, \citenamefont {Zhu}, \citenamefont {Hwang},\ and\
  \citenamefont {Yang}}]{Cheng2020}%
  \BibitemOpen
  \bibfield  {author} {\bibinfo {author} {\bibfnamefont {Yang}\ \bibnamefont
  {Cheng}}, \bibinfo {author} {\bibfnamefont {Sisheng}\ \bibnamefont {Yu}},
  \bibinfo {author} {\bibfnamefont {Menglin}\ \bibnamefont {Zhu}}, \bibinfo
  {author} {\bibfnamefont {Jinwoo}\ \bibnamefont {Hwang}}, \ and\ \bibinfo
  {author} {\bibfnamefont {Fengyuan}\ \bibnamefont {Yang}},\ }\bibfield
  {title} {\enquote {\bibinfo {title} {{Electrical Switching of Tristate
  Antiferromagnetic N\'eel Order in
  $\ensuremath{\alpha}\text{\ensuremath{-}}{\mathrm{Fe}}_{2}{\mathrm{O}}_{3}$
  Epitaxial Films}},}\ }\href {\doibase 10.1103/PhysRevLett.124.027202}
  {\bibfield  {journal} {\bibinfo  {journal} {Phys. Rev. Lett.}\ }\textbf
  {\bibinfo {volume} {124}},\ \bibinfo {pages} {027202} (\bibinfo {year}
  {2020})}\BibitemShut {NoStop}%
\bibitem [{\citenamefont {Zhang}\ \emph {et~al.}(2019)\citenamefont {Zhang},
  \citenamefont {Finley}, \citenamefont {Safi},\ and\ \citenamefont
  {Liu}}]{LuqiaoPRL2020}%
  \BibitemOpen
  \bibfield  {author} {\bibinfo {author} {\bibfnamefont {Pengxiang}\
  \bibnamefont {Zhang}}, \bibinfo {author} {\bibfnamefont {Joseph}\
  \bibnamefont {Finley}}, \bibinfo {author} {\bibfnamefont {Taqiyyah}\
  \bibnamefont {Safi}}, \ and\ \bibinfo {author} {\bibfnamefont {Luqiao}\
  \bibnamefont {Liu}},\ }\bibfield  {title} {\enquote {\bibinfo {title}
  {{Quantitative Study on Current-Induced Effect in an Antiferromagnet
  Insulator/Pt Bilayer Film}},}\ }\href {\doibase
  10.1103/PhysRevLett.123.247206} {\bibfield  {journal} {\bibinfo  {journal}
  {Phys. Rev. Lett.}\ }\textbf {\bibinfo {volume} {123}},\ \bibinfo {pages}
  {247206} (\bibinfo {year} {2019})}\BibitemShut {NoStop}%
\bibitem [{\citenamefont {Moriyama}\ \emph {et~al.}(2018)\citenamefont
  {Moriyama}, \citenamefont {Oda}, \citenamefont {Ohkochi}, \citenamefont
  {Kimata},\ and\ \citenamefont {Ono}}]{Moriyama2018}%
  \BibitemOpen
  \bibfield  {author} {\bibinfo {author} {\bibfnamefont {Takahiro}\
  \bibnamefont {Moriyama}}, \bibinfo {author} {\bibfnamefont {Kent}\
  \bibnamefont {Oda}}, \bibinfo {author} {\bibfnamefont {Takuo}\ \bibnamefont
  {Ohkochi}}, \bibinfo {author} {\bibfnamefont {Motoi}\ \bibnamefont {Kimata}},
  \ and\ \bibinfo {author} {\bibfnamefont {Teruo}\ \bibnamefont {Ono}},\
  }\bibfield  {title} {\enquote {\bibinfo {title} {Spin torque control of
  antiferromagnetic moments in {NiO}},}\ }\href
  {https://doi.org/10.1038/s41598-018-32508-w} {\bibfield  {journal} {\bibinfo
  {journal} {Scientific Reports}\ }\textbf {\bibinfo {volume} {8}},\ \bibinfo
  {pages} {14167} (\bibinfo {year} {2018})}\BibitemShut {NoStop}%
\bibitem [{\citenamefont {{Baldrati, L. and Gomonay, O. and Ross, A. and
  Filianina, M. and Lebrun, R. and Ramos, R. and Leveille, C. and Fuhrmann, F.
  and Forrest, T. R. and Maccherozzi, F. and Valencia, S. and Kronast, F. and
  Saitoh, E. and Sinova, J. and Kl\"aui, M.}}(2019)}]{Baldrati2019}%
  \BibitemOpen
  \bibfield  {author} {\bibinfo {author} {\bibnamefont {{Baldrati, L. and
  Gomonay, O. and Ross, A. and Filianina, M. and Lebrun, R. and Ramos, R. and
  Leveille, C. and Fuhrmann, F. and Forrest, T. R. and Maccherozzi, F. and
  Valencia, S. and Kronast, F. and Saitoh, E. and Sinova, J. and Kl\"aui,
  M.}}},\ }\bibfield  {title} {\enquote {\bibinfo {title} {{Mechanism of N\'eel
  Order Switching in Antiferromagnetic Thin Films Revealed by Magnetotransport
  and Direct Imaging}},}\ }\href {\doibase 10.1103/PhysRevLett.123.177201}
  {\bibfield  {journal} {\bibinfo  {journal} {Phys. Rev. Lett.}\ }\textbf
  {\bibinfo {volume} {123}},\ \bibinfo {pages} {177201} (\bibinfo {year}
  {2019})}\BibitemShut {NoStop}%
\bibitem [{\citenamefont {Gray}\ \emph {et~al.}(2019)\citenamefont {Gray},
  \citenamefont {Moriyama}, \citenamefont {Sivadas}, \citenamefont {Stiehl},
  \citenamefont {Heron}, \citenamefont {Need}, \citenamefont {Kirby},
  \citenamefont {Low}, \citenamefont {Nowack}, \citenamefont {Schlom},
  \citenamefont {Ralph}, \citenamefont {Ono},\ and\ \citenamefont
  {Fuchs}}]{Gray2019}%
  \BibitemOpen
  \bibfield  {author} {\bibinfo {author} {\bibfnamefont {Isaiah}\ \bibnamefont
  {Gray}}, \bibinfo {author} {\bibfnamefont {Takahiro}\ \bibnamefont
  {Moriyama}}, \bibinfo {author} {\bibfnamefont {Nikhil}\ \bibnamefont
  {Sivadas}}, \bibinfo {author} {\bibfnamefont {Gregory~M.}\ \bibnamefont
  {Stiehl}}, \bibinfo {author} {\bibfnamefont {John~T.}\ \bibnamefont {Heron}},
  \bibinfo {author} {\bibfnamefont {Ryan}\ \bibnamefont {Need}}, \bibinfo
  {author} {\bibfnamefont {Brian~J.}\ \bibnamefont {Kirby}}, \bibinfo {author}
  {\bibfnamefont {David~H.}\ \bibnamefont {Low}}, \bibinfo {author}
  {\bibfnamefont {Katja~C.}\ \bibnamefont {Nowack}}, \bibinfo {author}
  {\bibfnamefont {Darrell~G.}\ \bibnamefont {Schlom}}, \bibinfo {author}
  {\bibfnamefont {Daniel~C.}\ \bibnamefont {Ralph}}, \bibinfo {author}
  {\bibfnamefont {Teruo}\ \bibnamefont {Ono}}, \ and\ \bibinfo {author}
  {\bibfnamefont {Gregory~D.}\ \bibnamefont {Fuchs}},\ }\bibfield  {title}
  {\enquote {\bibinfo {title} {{Spin Seebeck Imaging of Spin-Torque Switching
  in Antiferromagnetic $\mathrm{Pt}/\mathrm{NiO}$ Heterostructures}},}\ }\href
  {\doibase 10.1103/PhysRevX.9.041016} {\bibfield  {journal} {\bibinfo
  {journal} {Phys. Rev. X}\ }\textbf {\bibinfo {volume} {9}},\ \bibinfo {pages}
  {041016} (\bibinfo {year} {2019})}\BibitemShut {NoStop}%
\bibitem [{\citenamefont {Chiang}\ \emph {et~al.}(2019)\citenamefont {Chiang},
  \citenamefont {Huang}, \citenamefont {Qu}, \citenamefont {Wu},\ and\
  \citenamefont {Chien}}]{Chiang2019}%
  \BibitemOpen
  \bibfield  {author} {\bibinfo {author} {\bibfnamefont {C.~C.}\ \bibnamefont
  {Chiang}}, \bibinfo {author} {\bibfnamefont {S.~Y.}\ \bibnamefont {Huang}},
  \bibinfo {author} {\bibfnamefont {D.}~\bibnamefont {Qu}}, \bibinfo {author}
  {\bibfnamefont {P.~H.}\ \bibnamefont {Wu}}, \ and\ \bibinfo {author}
  {\bibfnamefont {C.~L.}\ \bibnamefont {Chien}},\ }\bibfield  {title} {\enquote
  {\bibinfo {title} {{Absence of Evidence of Electrical Switching of the
  Antiferromagnetic N\'eel Vector}},}\ }\href {\doibase
  10.1103/PhysRevLett.123.227203} {\bibfield  {journal} {\bibinfo  {journal}
  {Phys. Rev. Lett.}\ }\textbf {\bibinfo {volume} {123}},\ \bibinfo {pages}
  {227203} (\bibinfo {year} {2019})}\BibitemShut {NoStop}%
\bibitem [{\citenamefont {Churikova}\ \emph {et~al.}(2020)\citenamefont
  {Churikova}, \citenamefont {Bono}, \citenamefont {Neltner}, \citenamefont
  {Wittmann}, \citenamefont {Scipioni}, \citenamefont {Shepard}, \citenamefont
  {Newhouse-Illige}, \citenamefont {Greer},\ and\ \citenamefont
  {Beach}}]{Churikova2020}%
  \BibitemOpen
  \bibfield  {author} {\bibinfo {author} {\bibfnamefont {A.}~\bibnamefont
  {Churikova}}, \bibinfo {author} {\bibfnamefont {D.}~\bibnamefont {Bono}},
  \bibinfo {author} {\bibfnamefont {B.}~\bibnamefont {Neltner}}, \bibinfo
  {author} {\bibfnamefont {A.}~\bibnamefont {Wittmann}}, \bibinfo {author}
  {\bibfnamefont {L.}~\bibnamefont {Scipioni}}, \bibinfo {author}
  {\bibfnamefont {A.}~\bibnamefont {Shepard}}, \bibinfo {author} {\bibfnamefont
  {T.}~\bibnamefont {Newhouse-Illige}}, \bibinfo {author} {\bibfnamefont
  {J.}~\bibnamefont {Greer}}, \ and\ \bibinfo {author} {\bibfnamefont
  {G.~S.~D.}\ \bibnamefont {Beach}},\ }\bibfield  {title} {\enquote {\bibinfo
  {title} {{Non-magnetic origin of spin Hall magnetoresistance-like signals in
  Pt films and epitaxial NiO/Pt bilayers}},}\ }\href {\doibase
  10.1063/1.5134814} {\bibfield  {journal} {\bibinfo  {journal} {Applied
  Physics Letters}\ }\textbf {\bibinfo {volume} {116}},\ \bibinfo {pages}
  {022410} (\bibinfo {year} {2020})}\BibitemShut {NoStop}%
\bibitem [{\citenamefont {Doran}\ \emph {et~al.}(2012)\citenamefont {Doran},
  \citenamefont {Church}, \citenamefont {Miller}, \citenamefont {Morrison},
  \citenamefont {Young},\ and\ \citenamefont {Scholl}}]{DORAN2012340}%
  \BibitemOpen
  \bibfield  {author} {\bibinfo {author} {\bibfnamefont {Andrew}\ \bibnamefont
  {Doran}}, \bibinfo {author} {\bibfnamefont {Matthew}\ \bibnamefont {Church}},
  \bibinfo {author} {\bibfnamefont {Tom}\ \bibnamefont {Miller}}, \bibinfo
  {author} {\bibfnamefont {Greg}\ \bibnamefont {Morrison}}, \bibinfo {author}
  {\bibfnamefont {Anthony~T.}\ \bibnamefont {Young}}, \ and\ \bibinfo {author}
  {\bibfnamefont {Andreas}\ \bibnamefont {Scholl}},\ }\bibfield  {title}
  {\enquote {\bibinfo {title} {{Cryogenic PEEM at the Advanced Light
  Source}},}\ }\href {\doibase https://doi.org/10.1016/j.elspec.2012.05.005}
  {\bibfield  {journal} {\bibinfo  {journal} {Journal of Electron Spectroscopy
  and Related Phenomena}\ }\textbf {\bibinfo {volume} {185}},\ \bibinfo {pages}
  {340} (\bibinfo {year} {2012})}\BibitemShut {NoStop}%
\bibitem [{\citenamefont {Morrish}(1995)}]{Morrish1995}%
  \BibitemOpen
  \bibfield  {author} {\bibinfo {author} {\bibfnamefont {A.~H.}\ \bibnamefont
  {Morrish}},\ }\href {\doibase 10.1142/2518} {\emph {\bibinfo {title} {Canted
  Antiferromagnetism: Hematite}}}\ (\bibinfo  {publisher} {{World
  Scientific}},\ \bibinfo {year} {1995})\BibitemShut {NoStop}%
\bibitem [{\citenamefont {Meer}\ \emph {et~al.}(2020)\citenamefont {Meer},
  \citenamefont {Schreiber}, \citenamefont {Schmitt}, \citenamefont {Ramos},
  \citenamefont {Saitoh}, \citenamefont {Gomonay}, \citenamefont {Sinova},
  \citenamefont {Baldrati},\ and\ \citenamefont {Kl\"aui}}]{meer2020direct}%
  \BibitemOpen
  \bibfield  {author} {\bibinfo {author} {\bibfnamefont {H.}~\bibnamefont
  {Meer}}, \bibinfo {author} {\bibfnamefont {F.}~\bibnamefont {Schreiber}},
  \bibinfo {author} {\bibfnamefont {C.}~\bibnamefont {Schmitt}}, \bibinfo
  {author} {\bibfnamefont {R.}~\bibnamefont {Ramos}}, \bibinfo {author}
  {\bibfnamefont {E.}~\bibnamefont {Saitoh}}, \bibinfo {author} {\bibfnamefont
  {O.}~\bibnamefont {Gomonay}}, \bibinfo {author} {\bibfnamefont
  {J.}~\bibnamefont {Sinova}}, \bibinfo {author} {\bibfnamefont
  {L.}~\bibnamefont {Baldrati}}, \ and\ \bibinfo {author} {\bibfnamefont
  {M.}~\bibnamefont {Kl\"aui}},\ }\href@noop {} {\enquote {\bibinfo {title}
  {Direct imaging of current-induced antiferromagnetic switching revealing a
  pure thermomagnetoelastic switching mechanism},}\ } (\bibinfo {year}
  {2020}),\ \Eprint {http://arxiv.org/abs/2008.05219} {arXiv:2008.05219
  [cond-mat.mtrl-sci]} \BibitemShut {NoStop}%
\end{thebibliography}
\end{document}